# Microstructure, Surface Plasmon, Magneto-optic Surface Plasmon, and Sensitivity Properties of Magneto-plasmonic Co/Au Multilayers

**Conrad Rizal** [1,2], **Senior Member IEEE**

[1] Department of Electrical Engineering & Computer Science, York University, Toronto, ON, Canada

[2] GEM Systems Inc, Markham, ON, Canada

crizal@yorku.ca; c.s.rizal@ieee.org

**Abstract:** Microstructure properties of [Co 1. 2 nm / Au 2.0 nm] × N = 20 multilayers prepared using dc-magnetron sputtering is reported using X-ray reflectivity (XRR) and X-ray diffraction (XRD) analysis. XRR profiles of these multilayers showed excellent bilayer periodicity. The XRD spectra displayed Co layer thickness-dependent properties. However, annealing increased lateral tensile strain decreased compressive strain along the normal substrate direction. While surface roughness, crystallite grain size, and strain are affected by Co layer thickness, microstructure and periodicity are dominated by fcc-Au (111). Surface plasmon resonance (SPR) and magneto-optic SPR study of an optimized Co/Au multilayer in *Kretschmann* configuration for air-He and water-Methanol media showed an enhanced sensitivity by over 4X when excited at $\lambda = 785$ nm as compared to the conventional SPR configuration when excited at $\lambda = 632.8$ nm. This enhancement in MOSPR sensitivity means the detection limit of this class of transducers can be substantially improved by tuning Co/Au layer thickness, wavelength and incident angle of optical radiation.

## 1.0 Introduction

Artificially produced ferromagnetic (FM) nano-scale multilayers (MLs) are of significant scientific interest due to their remarkable magnetic properties, X-ray reflection (XRR) and X-ray diffraction (XRD) patterns and due to their application in advanced technologies such as magnetic recording, sensing, etc., [1, 2]. Such films offer a unique combination of properties such as transparency in X-ray domain, magneto-optical effects [3], tuning of surface plasmon [4, 5], good electrical conductivity for specific compositions and thicknesses, and broadband reflectivity for relatively thinner film layers [6].

Nano-scale FM-MLs are often deposited on glass substrates to create high-quality broadband mirrored surfaces. These find application in reflecting telescopes as alternatives to lens arrays to reduce image aberration. These MLs also find applications in X-ray optics [7], medical diagnostics [8], gamma-ray optics [9], and hard X-ray astronomy [10], where the properties of these structures are strongly affected by inter-diffusion of atoms from one layer to another, their interfacial structure, and composition.

The thickness of these MLs is usually expressed as [Co ($t_{Co}$)/Au ($t_{Au}$)] × N = 20, where $t_{Co}$ and $t_{Au}$ are Co and Au layer thicknesses, respectively, and N is the layer periodicity. Au and Co are immiscible metals, meaning they show a high lattice mismatch of about 14 %, leading to strong interface roughness in the as-deposited state [11-14]. However, this surface roughness and lattice mismatch can be significantly improved if these multilayers are adequately annealed under controlled conditions, and by doing so their optical, magnetic, and structural, and magneto-optic properties can be changed and tuned substantially [15-20].

It is a well-established fact that the structure of nano-scale MLs is strongly related to their deposition conditions, layer thickness [21], and strain state [22]. It has been demonstrated that an additional in-plane anisotropy can be induced in these nano-scaled MLs if oblique angle deposition and magnetic annealing techniques are employed [23, 24].

From the prospective of designing more efficient magneto-optical devices, and in particular, the surge of interest in these structures springs from their usefulness in excitation and tuning of plasmonic resonance, as these offer the possibility of combining plasmonic properties of the diamagnetic, in this case Au layer, with the magneto-optic effect of ferromagnetic layer, in this case Co [25]. Specifically, this combination of alternating diamagnetic and ferromagnetic layers allows tuning of magneto-optical (MO) Kerr effects in a plasmonic configuration for sensing and other applications [4, 26, 27]. Thin film multilayers consisting of Co/Au and photonic crystals in the periodic system have been also employed successfully as a sensitive sensor [4].

The study of structure of magnetic MLs in nano-scale is important as their optical and magnetic properties depend on layer roughness, grain strain and the periodicity as opposed to their composition. It is also important for the understanding of the physical origin of perpendicular magnetic anisotropy (PMA), which is necessary for the optimization of magnetic recording media [14] and in tuning surface plasmon excitation [25].

Characteristically, thin film MLs prepared in nano-scale show layer periodicity at length scale that is comparable to or larger than in bulk material [22]. XRD and XRR techniques are extensively employed to investigate structure dependent properties of these kinds of MLs as the former offers nano-scale structural information such as crystallite size, lattice strain, deformational lattice stress, and deformation energy density, etc. In contrast, the latter offers information on individual layer/multilayer periodicity and surface or inter-layer roughness, and all these parameters are extracted from XRR and XRD profiles.

By fitting the data from the specularly-reflected X-rays into *GenX*, a parametric model of the MLs thin films can be constructed [28]. However, the data fitting requires important material characteristics such as electron density, $\rho_e$, thickness, $d$, of each layer, and surface roughness $\sigma$. In the case of MLs consisting of atomic-scale dimensions, the $d$ is obtained from XRR plots using Bragg's law as $m \times \lambda = 2d \sin\theta$, where $m$ is the order of reflection. ML periodicity is proportional to $2\pi/d$, whereas the amplitude is proportional to the density difference [22]. For detail on these properties, see Refs.[1].

In the past we have investigated 3-d transition metal based MLs using variety of deposition methods, namely, pulse current deposition [29], rf-magnetron sputtering [1], and oblique angle and normal e-beam evaporation [24], and have reported their magnetoresistance and magnetic anisotropy effect. Depending on preparation methods, these MLs exhibited in-plane and perpendicular magnetic anisotropy which can be controlled by tuning structure, layer thickness, and annealing conditions as suggested by their XRR and XRD profiles [24]. To understand their structure in detail, using experimental XRR results, we also used *GenX* fitting method that allowed extraction of multilayer information such as multilayer periodicity, intra-layer strain, interface roughness of multilayer structure, etc.,.

These FM MLs, also known as magneto-optical plasmonic MLs, are of significant scientific interest due to the following: (i) remarkable surface plasmon resonance (SPR) and magneto-optic SPR (MOSPR) properties, (ii) high sensitivity (definition of sensitivity is given in Section 2.2), (iv) real-time bio-detection that is possible with high accuracy [5] in the visible and near-IR region, and (v) possibility of developing new technologies in bio-sensing, bio-imaging and so on [4, 26, 30, 31]. However, a suitable optimized excitation configuration is needed to improve the performance of the sensor and to achieve the characteristics mentioned above.

In literature, three commonly used configurations for plasmon excitation exist: the *Kretschmann-Raether* configuration, similarly to shown in Ref. [2]. In this case, a thin metallic multilayer with the dielectric constant of Au and Co and magneto-optic constant of Co, $\varepsilon_{mCo}$ is deposited on the Ta buffer layer and glass substrate whose refractive index is the same as the prism. This sensing medium is a probing sample (air/He and water/Methanol, in the present case) is kept in contact with the multilayer (This is further described in Section 3). The second is the *Otto* configuration [32], in which the sensing medium is kept between the prism and thin film, and metal like gold (Au) or silver (Ag) for supporting the surface plasmon wave. In the Otto configuration, controlling the gap between the metal and the prism is cumbersome. And, the third is the *grating* configuration [33]. The surface plasmon propagates along the metal/dielectric interface in all cases.

The MO-SPR sensor is based on the excitation of the TM polarized (p-polarized) light along with the metal-dielectric interface, and the principle is based on Attenuated Total Reflection (ATR) coupling method [34]. Out of all three different configurations, this work reports on *Kretschmann* configuration. This configuration is comparatively easier to fabricate, and the substrate and buffer layer effect is minimal on the sensitivity.

Air pollution and environmental conditions vary on almost every time, scale, and pattern. Detecting these changes requires highly selective and sensitive bio-sensors. Moreover, in literature, most of the optimization of sensors is performed for an air medium [31]. Sensors optimized for air alone can not always be relevant for biosensing application as the effective refractive index of air is smaller than most bio-samples. In addition, it includes the weighted volumetric contributions of water and surface molecule functional groups, and these demands sensors optimized specifically for water-medium.

Several modulation approaches have been explored to enhance the sensitivity of the SPR based sensors. For example, Konopsky et al. [35], Slavik et al. [36], and Belotelov et al. [4] have proposed excitation of SPR that propagates ultra-long distances. The quality factor of sensors with ultra-long propagation distance is determined by the ratio of the real to the imaginary part of the propagation constant, implying that it acquires a higher value if surface plasmons sustain more extended propagation modes. In this case, ways to improve the real part of the dielectric constant of the sensing surface was explored, and proposed a magneto-plasmonic configuration in which the Au and Co layer thicknesses were optimized.

In this paper, we present detailed analysis of both the small angle X-ray reflectivity and large angle X-ray diffraction profiles of multilayered super-lattice of Co/Au in the as-deposited and annealed states and study possible influence of layer thickness and annealing on microstructure of these structures. We also investigate the SPR and MOSPR sensors in gas and liquid media using new SPR and MOSPR sensitivity metrics that benchmark the performance. Finally, we show that both the SPR and MO-SPR sensitivity are enhanced significantly when the operating wavelength is increased from 632.8 nm (visible) to 785 nm (near-IR). In all cases, the sensitivity of MO-SPR sensors is found to be superior to SPR sensors.

## 2.0 Theoretical formalism

### 2.1 X-ray Analysis and Diffuse Scattering

The concept of X-ray reflectivity is derived from classical optical and Fresnel theory. It is slightly modified when X-ray is used as a probe, and the complex refractive index, $n$, of the material in X-ray domain is given as [37]:

$$n = 1 - \delta - i\beta_1 \qquad (1)$$

where, $\delta = (\frac{r_0 \lambda^2}{2\pi}) N_A (\frac{\rho z}{A_w})$ and $\beta_1 = \frac{\mu \lambda}{4\pi}$. Here, $r_0 = 2.82 \times 10^{-13}$ cm is the classical electron radius, $N_A$ is the Avogadro number, $\rho$ is the density of the metal, $z$ is the atomic number, $A_w$ is the atomic weight of the element and $\mu$ is the linear absorption coefficient of the material at the particular wavelength. The quantity $N_A(\frac{\rho z}{A_w})$ in (1) is equal to the number of electrons per cm$^3$ for a particular element.

The reflectivity of X-ray from a thin film plane at the interface is given by Fresnel equation as $I(\theta) = |R_{1,2}|^2$, where subscripts 1, 2 refer to adjacent layers in the multilayers. The boundary between the 1 and 2 layers is assumed to be continuous for the tangential components of the electric field vector, and its derivative at the interface becomes a continuously differentiable function. The reflectivity of the ML stack is then obtained using Fresnel's equation at each successive interface iteratively [22].

Since the refractive index of all material in X-ray domain is less than unity, the probing beam experiences an oscillating effect due to the surface roughness of the ML, mainly to the existence of a smooth interface for one type of interface whereas rough for another type. In this case, we are dealing with two interfaces types: Co-Au and Au-Co.

As we assume the interface to be continuous from the thickness of 1.2 to 1.8 nm, it is obvious that the tangential component of the incident electric field vector is continuous at the interface, which in turn means its derivative is a continuous function, prompting the need for modification of Fresnel equation. The reflectivity of the ML stack is then obtained using successive application of the Fresnel's equation at each interface, as it is believed that the reflectivity at very low angle of incidence approaches to $\cong 1$, and in which case, the beam undergoes total external reflection as $n$ tends to be less then unity. In this case, the critical angle $\theta_c$ depends on $\delta$, and after approximation, it is defined as $\theta_c \approx \sqrt{2\delta}$. It, in turn, implies that with an increasing angle of incidence, the X-ray beams start penetrating inside the layer, and the reflectivity falls off at a rate defined by the absorption loss on the stack.

Typically, a reflectivity tail is observed for all the MLs as well, which is usually described by the oscillations of the X-rays due to the total internal reflection within the thickness of the film in case of a single layer sample. In contrast, the oscillations of the beam define it within the bilayer for ML samples. The specular X-ray reflectivity spectrum analysis can yield approximate information about the density and thickness of a thin multilayer [37].

However, the Fresnel's reflectivity needs to be modified for a rough surface such as the ones under consideration and it can be done using "*Debye-Wallar-like*" approach as in Refs. [38]. The reflectivity of a rough surface, $R_o$ is given as:

$$R_0 = R_u e^{\left(-\frac{v^2 \sigma^2}{2}\right)}. \qquad (2)$$

where $v$ is the momentum transfer factor, and it is given as $v = \frac{4\pi \sin\theta}{\lambda}$, and $R_u$ is the reflectivity of an identical smooth surface, and $\sigma$ is the root mean square, rms, the roughness of the surface. The *rms* roughness value is quantitatively equal to interface width $\sigma$ for interfaces and it, in turn, equals the sum of the interfacial roughness and interfacial diffusion [39].

In an ideal case, all crystals are finite and ideally extend infinitely in all directions. However, in practice, the lattice size of the deposited film can deviate (increase or decrease) significantly from its ideal state. This deviation from perfection is responsible for the shift in Bragg's angle position and broadening of the XRD peaks. Two important properties of crystals can be extracted from this broadening of Bragg peaks: crystallite size and lattice strain. Crystallite size measures the deviation from the coherently diffracting dimension/size, and it differs in pure crystal, polycrystalline aggregates, and amorphous material.

Lattice strain $\varepsilon$ is a measure of deviation in lattice constant $a$, arising from crystal imperfection due to lattice dislocation and size mismatch. Other sources of crystal imperfection include coherent stress, sintering stress, stacking faults, grain boundary, and defects [40]. Crystallite size and strain are well known for showing the strong effect on Bragg peaks differently. The manifestations are mainly in the form of change in peak width, intensity, and shift in peak position.

An unstrained grain with uniformly spaced, $a$ in equilibrium shows diffraction lines that are equally placed. For example, a diffraction line from a crystal plane under uniform tensile strain at a right angle to the reflecting plane is shifted to a lower Bragg angle as their inter-planer spacing $d$ becomes greater than $d_e$. Here, $d_e$ means inter-planer spacing at equilibrium condition. However, the intensity and peak width do not change in this case. Therefore This change is the basis for the estimation of micro-strain in the layer.

The crystal lattice bends when a strain is applied normal to it, and the spacing on the reflecting surface exceeds $d_e$ at the top whilst it is less than $d_e$ on the bottom side, and in the middle, it is $d_0$. The diffraction peak in this case shrinks and broadens proportionally in response to the compression on the substrate/bottom plane [41].

The peak broadening induced by $\varepsilon$ arises from crystal bending, imperfection, and distortions and is related to surface and interface roughness, which is given as $\varepsilon = \beta_s/4\tan\theta$ where $\beta_s$ is the breadth of the peak, which is estimated as full-width-half-maximum (FWHM) [41]. The $\beta_s$ which changes proportionally to grain size varies as $1/\cos\theta$, whereas the $\varepsilon$ varies as $1/\tan\theta$. The Bragg width contribution from the crystallite size is proportional to breadth $\beta$ of the peak.

Since both the crystalline size and $\varepsilon$ are related to peak broadening, separating one effect from another is essential. This is typically done using the Williamson-Hall approach, which assumes that the size and strain components are additive and need separation. This separation, D is achieved through combining Scherer's equation, $k\lambda/(\beta_c \cos\theta)$ with strain equation $\beta_s/\tan\theta$ as $\beta_{hkl} = \beta_c + \beta_s$, where $\beta_c$ and $\beta_s$ denote strain due to crystalline size and lattice strain, respectively, and k is a constant and usually it is assumed to be 0.97.

It means $\beta_{hkl} = k\lambda/D\cos\theta + 4\varepsilon \tan\theta$, which upon rearranging gives, $\beta_{hkl} \cos\theta = k\lambda/D + 4\varepsilon \sin\theta$. *Willimson-Hall* plot is a graph of $\beta_{hkl} \cos\theta$ versus $4 \sin\theta$, which gives a straight line and whose slope is equal to strain, $\varepsilon$ and whose y-intercept is equal to grain size [41]. Here, D is related to peak width and position as $D = k\lambda/(\beta_d \cos\theta)$ where $k$ is the shape factor and $\beta_d$ is the peak breadth at FWHM. A plot *of cosθ* versus *1/β_d* thus gives a slope which is equal to *D*, and the intercept gives them $\varepsilon$. This approach is also known as a uniformly deformable model. It is valid as long as all the material properties are independent of crystallographic directions, meaning it is valid as long as the MLs being considered are isotropic, uniform, and homogeneous. In this plot, negative slope signifies negative $\varepsilon$, representing shrinkage of lattice constant - corresponding to compressive $\varepsilon$, whereas positive slope suggests positive $\varepsilon$, indicating expansion of lattice constant- corresponding to tensile $\varepsilon$, for details, see Ref. [1].

For MLs consisted of bilayers in the atomic scale, correlation length, ξ, normal to the substrate plane can also be calculated using Scherer's formulas as ξ = kλ /β cosθ, where k is shape factor of the crystallites, and as stated above, approximately ≈ 0.97, β is the FWHM of the Bragg peak, λ is the wavelength of the X-ray beam [42].

Another model that can be used to analyze peak broadening is the uniform stress deformation model, which employs *Hook's law* in its generalized form to study the changes in peak width. It can be done if the stress (σ) - strain (ε) plot shows linear relationship as σ = Yε, where Y is Young's modulus. Combining Hook's equation with strain equation (ε=$β_s$/tanθ ) and $β_{hkl}$ = kλ/($β_d$ cos θ + $β_s$/tanθ gives $β_{hkl}$ cos θ = k λ/D + 4 σ sinθ / $Y_{hkl}$. The plot of $β_{hkl}$ cos θ versus 4 σ sinθ / $Y_{hkl}$) gives a straight line whose slope is equal to stress and whose Y-intercept gives to crystalline size.

## 2.2 SPR, MOSPR, and Sensitivity

The sensitivity of surface plasmon resonance (SPR) and magneto-optic SPR (MO-SPR) sensors has been defined in many different ways in characterizing plasmonic-based sensors, generally depending on the sensor response or mode of excitation [43, 44]. However, all these can lead to some confusion about how sensitivity should be defined, particularly when comparing sensors of different types [27, 30, 44, 45]. This paper reports new performance metrics that we have recently developed for gas and liquid media that would allow one to accurately quantify and compare the performance of all types of SPR and MO-SPR sensors in terms of the measured quantities as follows [2, 46]:

For SPR sensors, we define sensitivity as: $\left\{\left(\frac{R_p(A)-R_p(B)}{R_{p(A)m}}\right) \times 100\right\} \frac{1}{\Delta n}$ [% RIU$^{-1}$], where, $R_{p(A)m}$ is the magnitude of reflected intensity at an incident angle $\theta_m$ where $\theta_m$ is the incident angle at which the first derivative $\left|\frac{dR_p(\theta)}{d\theta}\right|$ is maximized, and RIU stands for refractive index unit. The difference $R_p(A) - R_p(B)$ is normalized by $R_{p(A)m}$. This is also an experimental condition that gives the best sensor performance. For gas media, A and B are air and Helium, whereas in liquid media, A and B are water and Methanol, and $\Delta n$ is the difference in refractive indices between air and He or water and Methanol.

For MOSPR sensors, we define sensitivity as: $\left\{\left(\frac{\Delta R_p(A)-\Delta R_p(B)}{\Delta R_{p(A)m}}\right) \times 100\right\} \frac{1}{\Delta n}$ [% RIU$^{-1}$], where $\Delta R_p(A)$ and $\Delta R_p(B)$ are the change in reflectivity due to modulating H field for air and He or water and Methanol medium, respectively at $\theta_m$. The difference $\Delta R_p(A) - \Delta R_p(B)$ is normalized by $\Delta R_{p(A)m}$ where $\Delta R_{p(A)m}$ is the maximum change at the first derivative $\left|\frac{d\Delta R_{p(A)}(\theta)}{d\theta(A)}\right|$. For air media, A and B are air and Helium gases, whereas in water media, A and B are water and Methanol, and $\Delta n$ is the difference in refractive indices between air and He or water and Methanol.

Unlike the sensitivity metrics reported in various literature in the past, our sensitivity metrics eliminate diverging sensor response since both $R_{p(A)m}$ and $\Delta n$ in the SPR case and $\Delta R_{p(A)m}$ and $\Delta n$ in the MOSPR case is never zero at the incident angle, $\theta_m$, near the resonance angle $\theta_{SPR}$, as $\theta_m$ is chosen where the slope of the curve is maximized. Moreover, using these sensitivity metrics, we can directly compare the performance of SPR sensors, where typically sensitivity refers to an angular shift of the resonance peak per RIU change in the measurand, with that of MO-SPR sensors where typically sensitivity refers to a change of intensity per RIU change.

## 3.0 Experimental Methods

Three samples of Co/Au multilayers with Au layer thickness, $t_{Au}$ = 2 nm and varying Co layer thickness, $t_{Co}$ have been deposited using rf-magnetron sputtering system onto glass slide BK-7 / Crystalline silicon (111) wafer as a substrate and the deposition was carried at an ambient temperature and at a vacuum pressure of $3 \times 10^{-8}$ Torr. A Ta buffer layer with 2 nm was deposited before the deposition of 2 nm Au layer. The multilayer samples were annealed at 400 $^0$C using a rapid thermal furnace (an AG Associates Heat Pulse 610) for 30 minutes and under a pressure P of $2 \times 10^{-6}$ Torr. For detail on deposition, see, Ref. [1].

The system vacuum was created using turbo-pump and backed by a rotary pump. The device contained two water-cooled cathodes with a circular substrate holder. A 1.5 kW power was used to supply rf-power to the cathodes. The substrate holder is also water-cooled to cancel out the temperature grown by the plasma during the sputtering process. A shutter was placed between the cathode and anode. To monitor layer thickness, quartz crystal monitors have also been used. In all cases, high purity Co, Au, and Ta targets were used. Various parameters used for the deposition are listed in Table 1.

| Table 1. Sample deposition parameters | | | | |
|---|---|---|---|---|
| Bilayer, N | Substrate | [b]$t_{Ta}$ (nm) | [c]$t_{Co}$ (nm) | [d]$t_{Au}$ (nm) |
| 20 | Glass / SiO$_2$ | 2.0 | 1.2 - 1.8 | 2.0 |
| [a] Multilayer periodicity, [b] Ta layer thicknesses, [c] Co and [d] Au layer thicknesses, respectively. | | | | |

A *Rigaku Smart Lab* diffractometer, with *Cu Kα* radiation of λ = 0.1542 nm was used to study low angle X-ray reflectivity (XRR) and high angle X-ray diffraction (XRD) to extract information on microstructure of the fabricated Co/Au multilayer samples. The low angle XRR method was used to estimate the multilayer thickness. Measured data were analyzed and plotted both XRR and XRD profiles using *KyPlot* - a data-analysis, graphing, and drawing program [47].

The magneto-optic effects were modeled using finite off-diagonal components of the complex permittivity tensor [25] and transfer-matrix formalism. An anti-reflection coated right-angled prism was chosen to allow the rotation of the prism without refraction of the incident and reflected light at the air/He-prism or water/Methanol-prism boundary by directing the TM polarized (p-polarized) optical radiation at the wavelengths of 632.8 nm and 785 nm. Additionally, a 2 nm thick Ta was used as a buffer layer. Although T$_a$ is lossy, it does not help with the sensor response, and most theoretical works have neglected its effect; the effect of T$_a$ was included in our calculation, as in practice, these adhesion layers are necessary for good fabrication outcome. For details on excitation configuration, interested readers are referred to Ref. [2]. The various dimensional and optical parameters (refractive indices, etc.) used in the calculations are given in Table 2.

**Table 2: Optical and geometrical parameters of sensor configurations at λ = 632.8 nm and 785 nm. ε$_{mo}$ denotes the magneto-optical component of the dielectric function, set to literature values under the influence of saturating H field [48].**

| Material | Thickness, t (nm) | Permittivity, ε ; ε$_{mo}$ at λ = 632.8 nm | Permittivity, ε; ε$_{mo}$ at λ = 785 nm |
|---|---|---|---|
| Substrate | 1300 | 2.2940 | 2.2813 |
| BK-7 Prism | - | 2.2940 | 2.2813 |
| Index-match Liquid | - | 2.2940 | 2.2813 |
| Au | 2 | -11.127+j1.3268 | -22.855+j1.4245 |
| Co | 1.2 | -12.475+j18.451; -0.65+j0.0005 | -16.493+j23.377; -0.85+j0.0006 |
| Ta | 2 | -31.101+12.005 | -50.104+j14.240 |
| He | - | 1.00006980 | 1.00006960 |
| Air | - | 1.0005530 | 1.0005502 |
| Methanol | - | 1.759 | 1.7296 |
| Water | - | 1.7734 | 1.7678 |

# 4.0 Results and discussion

## 4.1 XRR Analysis

Nano-scale single-layered films of Co or Au deposited on glass substrate are usually amorphous in the as-deposited state. These characteristically show continuously disordered, rough surfaces with local fluctuations in layer thickness mainly due to the roughness of the substrate. Annealing helps smooth out these defects and enhances crystallization.

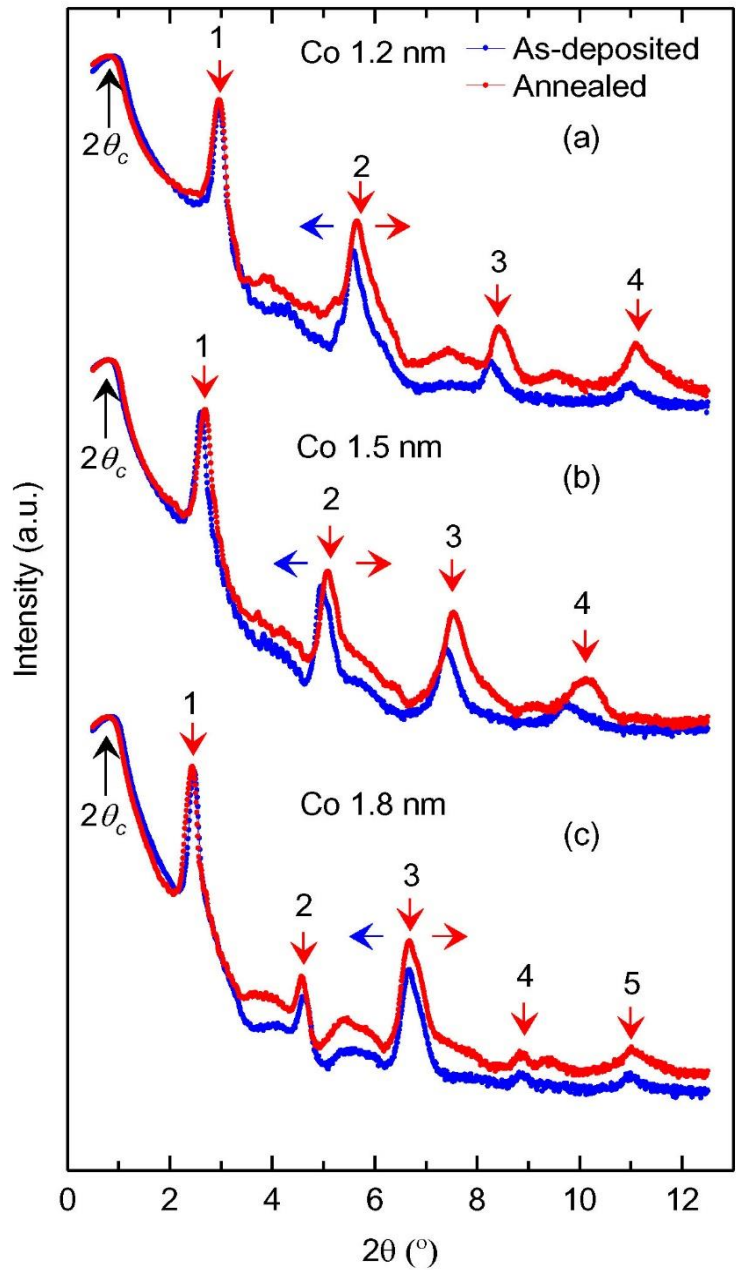

**Fig. 1:** Co layer thickness dependence of XRR profiles of as-deposited (blue circles/line) and annealed samples (blue circles/line) of Au/Co layers stacked as a multilayered structure with $t_{Co}$ = 1.2, 1.5, and 1.8 nm thickness. The curves below the critical angle, $2\theta \approx 0.8°$ is not shown here, since some systematic deviations are observed below this angle due to over-illumination of the relatively wide incident X-ray beam. The thick horizontal blue and red arrows indicate the direction of shift of the curves for as-deposited (left) and annealed (right) samples, respectively. Numbers 1 through 5 indicate the position of reflection peaks.

Co layer thickness-dependent XRR of Co/Au multilayers of as-deposited (denoted by blue curves) and annealed (denoted by red curves) with $t_{Co}$ = 1.2, 1.5 and 1.8 nm thickness are shown in Fig 1 (a-c). Both the curves show distinct reflectivity profiles corresponding to the average inter-planar distance of Co and Au. They show the signature of the continuous rough layer as suggested by the presence of monotonously decreasing reflection intensity which tails with the angle with increasing Bragg angle. The peak positions for the annealed samples shifted slightly towards a higher angle, meaning that the ML shrinks somewhat. On the other hand, the peak intensity increased with annealing, signifying improved interface (This is not explicitly visible in the normalized curves shown in Fig. 1).

The bilayer periodicity at low angle can be estimated using Bragg's law as $n\lambda = 2d\sin\theta$, where d in this case is bilayer periodicity as opposed to grain lattice spacing in XRD mode, which will be discussed in the next section. As shown in the figure, all the multilayers showed distinct Bragg's peaks corresponding to the bilayer periodicity. However, the higher order Bragg peaks, up to the 5th order in Fig 1(c) in XRR spectra, further indicate that, despite the rather small layer thickness comparable to atomic-scale dimension, all the samples studied here maintain good layer periodicity.

The broadening of the XRR peak with $t_{Co}$ suggests morphology variation and variation in lattice strain, as will be shown later in XRD study. A slight broadening of the Bragg peaks near the continuous-to-discontinuous regime can also be observed as the layer thickness changed, and this effect became more prominent for the samples with $t_{Co}$ = 1.5 nm. It is widely believed that this kind of change is due to the significant variation in morphology of the film with a change in Co layer thickness.

Our earlier paper reports the nominal, experimentally obtained parameters obtained from Bragg's law, and data fitting of both the as-deposited and annealed Co/Au multilayer with similar structures [1]. These are not discussed here

From the comparison of XRR profiles of samples with different thicknesses in Fig. 1, it can be seen that the interface diffusion also follows similar trends, as does interface width. As can be seen, as the Co layer thickness decreased, the slope of the curve increased, meaning physical roughness increased for all samples and vice versa. and it is found to be the case at both the interfaces. As far as layer density is concerned, it increased with layer thickness and with increasing density, the inter-diffusion of atoms between adjacent layers decreased profoundly. A similar type of observation was reported by *Dekadjevi et al.* [49] in the case of the Fe/Au ML system with a discontinuous low-density Fe layer deposited on a high-density Au layer.

The characteristics signature of bilayer periodicity is mainly defined and controlled by the relative difference in refractive indices of the two materials, in this case, Au and Co. It is in turn defined by the atomic density and number of electrons per atom contributing to the scattering of the probing beam (in this case, X-rays), which is also described as the scattering power of the material. Au with more electrons possesses stronger scattering power, meaning the property of bilayer in this case is defined by the reflections from the Au layer. The reflections from bilayers is also contributed by interface roughness [37].

Much like e-beam evaporated MLs of Au presented by us previously [24], the rf-sputter deposited Co/Au MLs also showed an increase in the amplitude of oscillation of *Kiessig fringes* with increasing film thickness. In contrast, the magnitude of the reflectivity decreased faster with increasing Bragg's angle, signifying that the top surface roughness of the films increased proportionally with increasing the thickness of the Co layer,. See our earlier paper [1] again.

**4.2 XRD Analysis**

The XRD spectra of a single Au layer of 40 nm thickness, measured in the range of Bragg angle, $2\theta = 30$ to $90^0$ is given in Fig. 2 (a). The Au layer was deposited over an amorphous glass substrate. As shown, a single Au layer exhibited a dominant Bragg peak at $38^0$, a signature of fcc-Au (111), signifying pre-dominant crystallization of Au into fcc-shaped crystallites. Besides Au (111), other crystallographic phases of Au, such as Au (200), Au (220), Au (331), and Au (222) are also observed.

As shown in Fig. 2(b), the stacks of Au/Co ML super-lattice, on the other hand, showed only two dominant Bragg peaks and two satellite peaks. The two intense peaks appeared at $37^0$ and $40^0$ whereas the small satellite peaks appeared at $35^0$ and $43^0$. It is assumed that the intense peak that occurred at $37^0$ is due to Au (111) fcc-phase, which is shifted towards a lower angle, implying that the crystallite size increased in MLs instead of a single layer.

Another exciting aspect of these ML Bragg peaks is that they tend to show a pattern in which they appear in a repeating order, and in this case, with a repeating pattern at the interval of about $3^0$. This repeating pattern is thought to be due to the high periodicity of the bilayers, as a result of the periodic refractive index difference between each Co/Au layer. These four peaks appear at the same interval of $3^0$ most certainly signifies the presence of a periodic system that dominates the X-ray diffraction property of these MLs instead of layer composition and type of metal making these layers. The reflection peaks due to Co are not observed for the multilayer here.

It is also to be noted that the intensity of the Bragg's peaks due to fcc-Au is about five-fold higher in case of the Superlattice MLs (it is not explicitly observable in the normalized curves) as compared to the single layer Au- which in turn indicates that the Au layer is periodic in the latter case. On the other hand, the intensity of peak at around $45^0$ pertaining to Au (200) and Au (311) decreased dramatically for the MLs, suggesting the fact that the geometrical growth of the crystals in each layer is dominated by the Au (111) phase in these ML structures. The fact that the full-width at half-maximum, FWHM, of the Bragg peak of these structures, is smaller than the FWHM of the single-layer suggests that the Au crystals in ML reduced significantly compared to Au single layer and vice versa. Contraction of grain size in these MLs, as opposed to single layer, suggests that crystallite grains are strained in the former compared to the latter, as strain $\varepsilon$ varies proportionally to the variation of $tan\theta$.

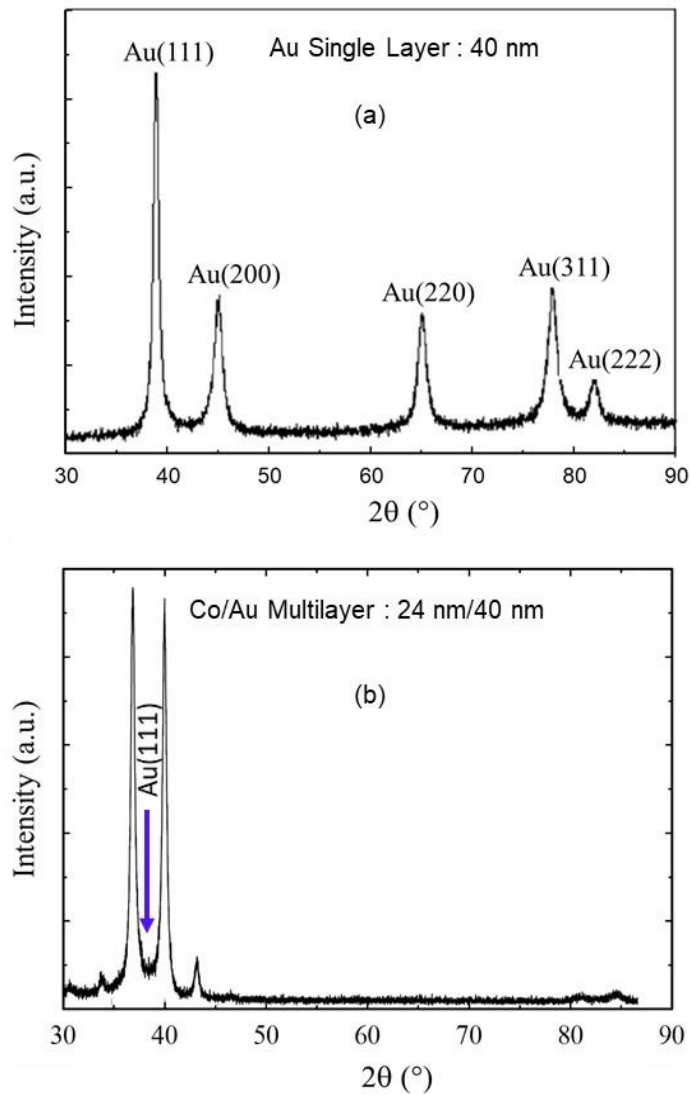

**Fig. 2.** XRD profile of (a) as-deposited Au single layer of $t_{Au}$ = 40 nm and (b) Co/Au multilayers with $t_{Co}$ = 1.2 and $t_{Au}$ = 2.0 nm with 20 repeats.

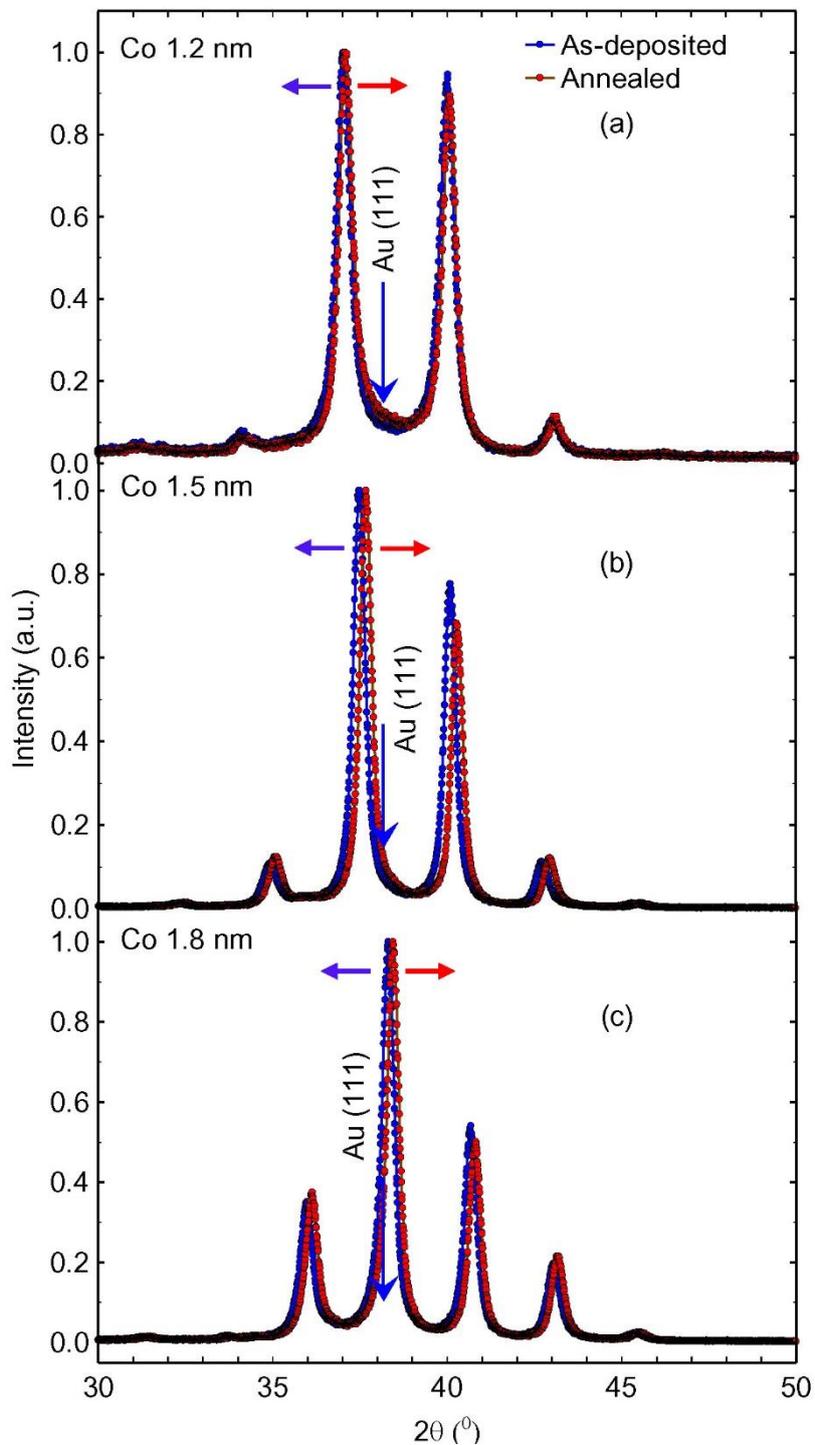

**Figure 3.** Comparison of XRD profiles of as-deposited and annealed samples. The thick blue vertical arrows indicate the ideal position of Au<111>. The thick horizontal red arrows indicate the direction of the shift of the curves for as-deposited (left) and annealed (right) samples, respectively.

Comparison of XRD spectra of as-deposited and annealed samples with $t_{Co}$ = 1.2, 1.5, and 1.8 nm is given in Fig 3 (a-c). As shown, slight changes in Bragg peak intensity, position, and full width at half maximum (FWHM) are observed, implying a change in grain size and strain with annealing. Annealed samples showed increase in peak intensity compared to as-deposited samples (this is not explicitly visible in the normalized XRD profile), implying that grain crystallinity enhanced. A slight red-shift in peak position accompanied by a change in peak width means the grains change crystal strain. In this case, the FWHM of the peak widened slightly with annealing, suggesting that tensile strain increased along the lateral direction, whereas the compressive strain along the vertical axis decreased slightly. Similar trends have been observed for $t_{Co}$ = 1.5 and 1.8 nm.

The structural coherence length, $\xi$, which is the distance over which the atomic positions are quantitatively correlated, was estimated from the FWHM, $\beta$, of XRD peaks using Scherer's relations as: $\xi = 0.9 \times \lambda / (\beta \cos \theta_x)$, where $\theta_x$ is the Bragg angle. In many multilayer systems, the stacks' periodicity usually repeats only for a few layers due to structural disorder. However, in the present case, $\xi$ is found to be repeated up to 8 atomic planes, estimated normal to the substrate surface plane for our annealed samples. Observing the strong satellite peaks in both the as-deposited and annealed multilayers suggests that the multilayers have strong structural coherence of atomic planes [11,21].

### 4.3 SPR versus MOSPR

Figure 4 (a-d) show SPR and MOSPR profiles: (a) Rp vs θ, (b) $\Delta R_p$ [$R_{p(H=H)} - R_{P(H=0)}$] vs θ, (c) SPR sensitivity and (d) MO-SPR sensitivity for the [Co 1.2 nm /Au 2.0 nm] × N = 12 multilayers. The dielectric media is varied from (a) air to He and (b) water to Methanol and the optical response is calculated for TM-polarized light at $\lambda$ = 632.8 nm.

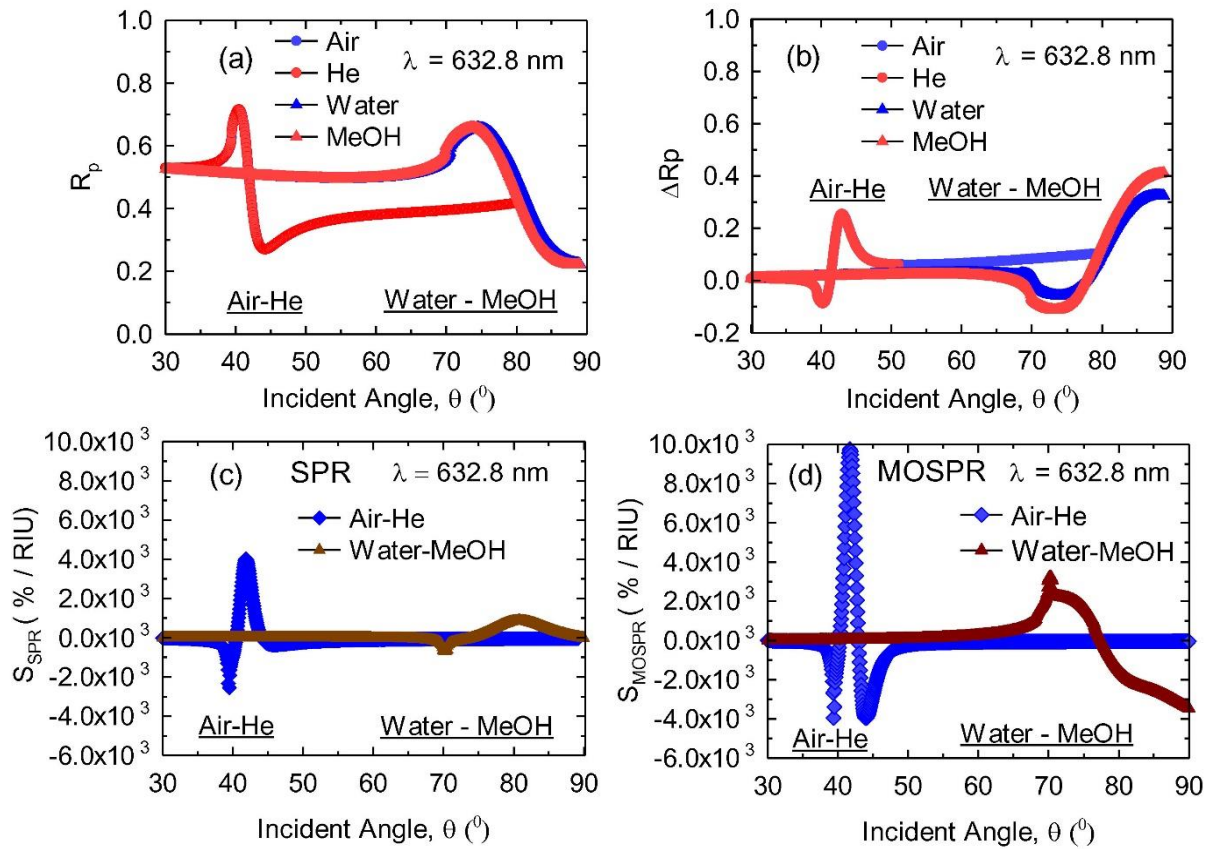

**Fig 4. (a) $R_p$ vs incident angle, θ (b) $\Delta R_p$ vs θ, (c) SPR sensitivity, and (d) MOSPR sensitivity for air-He and water – Methanol at $\lambda$ = 632.8 nm. The thickness of Co is kept at 1.2 nm and Au at 2.0 nm.**

As shown in Fig. 4, both the minimum reflectivity and peak positions are media dependent, and they significantly vary between air-He and water-Methanol media.

In the case λ = 632.8 nm, the minimum reflectance observed are Rp ≈ 0.22 for air-He and ≈ 0.18 for water-Methanol media, respectively.

As shown for λ = 632.8 nm, the multilayer has an SPR effect of $4.0 \times 10^{03}$ % / RIU (defined as % change in detected signal per unit change in the dielectric medium of the sensor) and MOSPR effect of $1.0 \times 10^{04}$ % / RIU for air-He medium. For the water-Methanol medium, the multilayer shows an SPR effect of $1.5 \times 10^{03}$ % / RIU and a MOSPR effect of $2.0 \times 10^{03}$ % / RIU, respectively; RIU stands for refractive index unit.

Figure 5 (a-d) show SPR and MOSPR profiles: (a) Rp vs θ, (b) ΔRp vs θ, (c) SPR sensitivity and (d) MO-SPR sensitivity for [Co 1.2 nm / Au 2 nm] × N = 12 multilayers, with air-He and water-Methanol media, calculated again for TM-polarized light at λ = 785 nm. In the case of λ = 785 nm, the minimum reflectance observed are Rp ≈ 0.23 for air-He and 0.19 for water-Methanol, respectively. The multilayer showed an SPR effect of $8.0 \times 10^{03}$ % / RIU and MOSPR effect of $3.0 \times 10^{04}$ % / RIU air-He media. For the water-Methanol media, the multilayer has an SPR effect of $4.0 \times 10^{03}$ % / RIU and a MOSPR effect of $1.0 \times 10^{04}$ % / RIU, respectively.

At both wavelengths of operation, the SPR and MOSPR profiles are sharper for the air-He media as compared to the water-Methanol media.

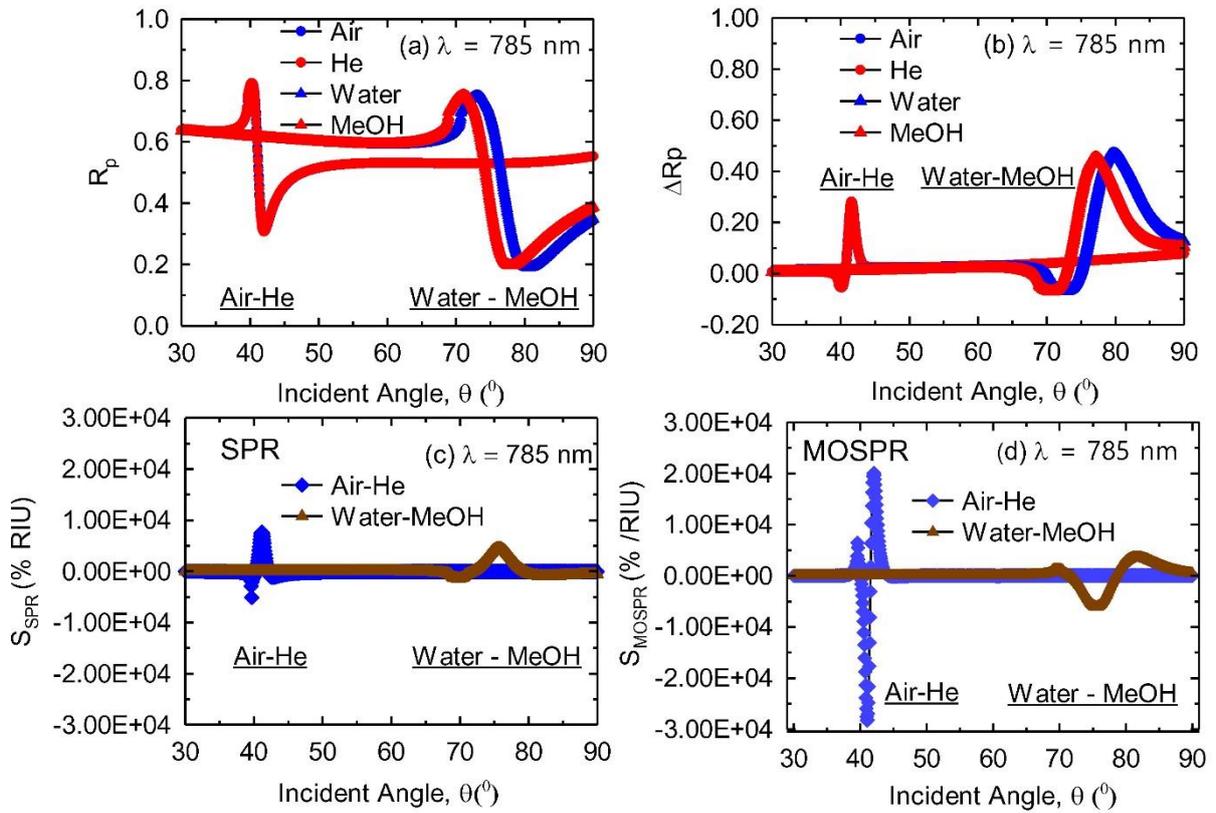

**Fig. 5. (a) $R_p$ vs incident angle, θ (b) ΔRp vs θ, (c) SPR sensitivity, and (d) MOSPR sensitivity for air-He and water-Methanol media at (a) λ = 785 nm. Thickness of Co is kept at 1.2 nm and Au at 2.0 nm.at**

Sensitivity values were obtained from Figs. 4 and 5, and they are listed in Table 3. As shown, the MOSPR sensitivity at 632.8 nm is increased by 2.5X for air-He media and 2X for water-Methanol media. Similarly, the MOSPR sensitivity at 785 nm is increased by ≈ 4X for air-He media and ≈ 2.5X for water-Methanol media.

**Table 3. Comparison of SPR and MOSPR Sensitivity for air-He and Water-Methanol media at 632.8 and 785 nm.**

| Probing Material | SPR Sensitivity (% / RIU) | MOSPR Sensitivity (% / RIU) |
|---|---|---|
| **Excitation at λ = 632.8 nm** | | |
| air-He | 4.0E+03 | 10.0E+03 |
| water-Methanol | 1.5E+03 | 03.0E+03 |

| Excitation at λ = 785 nm | | |
|---|---|---|
| air-He | 8.0E+03 | 3.0E+04 |
| water-Methanol | 4.0E+03 | 1.2E+04 |

In both the cases, we found that the MO-SPR sensitivity was enhanced by up to 4× concerning the SPR sensitivity in both air-He and water-Methanol media studied here. The sensitivity of the MOSPR based sensors for a liquid medium is also comparable to that of the MOSPR sensors for gas medium, which targets common biosensing platforms.

The marked differences in SPR and MOSPR sensitivities obtained for air-He versus water-Methanol media highlights that this class of sensors require layer thickness optimization and performance evaluation according to the clinical applications.

**5.0 Conclusions**

We reported microstructure, surface plasmon resonance, magneto-optic surface plasmon resonance, and nano-scale Co/Au multilayers sensitivity studies. The low angle XRR profiles showed increasing surface roughness with Co layer thickness, and it improved with annealing temperature, and the morphology is dominated by fcc-Au (111). XRR profiles of these multilayers also showed excellent bilayer periodicity. In addition, the high angle XRD profiles displayed Co layer thickness-dependent properties.

We highlighted differences between the SPR responses in air and water media and benchmarked sensitivity metrics that permitted direct comparison of the performance of SPR and MOSPR sensors at the optimal operating point. Our design yielded over 4X magnitude improvements in sensitivity when excited at 785 nm as compared to the conventional SPR configuration, which is excited at λ = 632.8 nm. This enhancement in sensitivity means the detection limit of this class of transducers can be substantially improved by tuning the layer thickness, wavelength, and incident angle of optical radiation. The minimum reflectivity and peak positions are also wavelength dependent and they significantly vary for air and water media. Moreover, the sensitivity metrics reported in this paper avoid issues with diverging sensitivities.

The sensor responses in air medium are more robust and relevant for environmental and pollution studies. The sensor response in water medium is also found to be more substantial compared to the sensor response of conventional SPR sensors, and is highly relevant for bio-sensing that involves liquid media.

**Acknowledgement**

The author acknowledges the financial support from the Centre for Memory and Recording Research (CMRR) at UC San Diego, the USA, and Natural Sciences and Engineering Research Council (NSERC), and MITACS Inc. of Canada. The author thanks Dr. Boris B. Niraula (Seed NanoTech International Inc., Canada) for providing valuable feedback on the microstructure study at the early manuscript preparation stage.